\documentclass{ws-ijmpd}
\begin{document}

\markboth{Chengwu Zhang, Lixin Xu, Yongli Ping and Hongya
Liu}{Reconstruction of $5D$ Cosmological Model From Recent
Observations}

\catchline{}{}{}{}{}

\title{Reconstruction of $5D$ Cosmological Models From Recent
Observations}

\author{Chengwu Zhang\footnote{zhangcw@student.dlut.edu.cn},
Lixin Xu, Yongli Ping and Hongya Liu\footnote{hyliu@dlut.edu.cn}}

\address{School of Physics and Optoelectronic Technology, Dalian
University of Technology, Dalian,116024, P.R. China}

\maketitle

\begin{history}
\received{Day Month Year} \revised{Day Month Year}
\end{history}

\begin{abstract}
We use a parameterized equation of state (EOS) of dark energy to a
$5D$ Ricci-flat cosmological solution and suppose the universe
contains two major components: dark matter and dark energy. Using
the recent observational datasets: the latest 182 type Ia Supernovae
Gold data, the 3-year WMAP CMB shift parameter and the SDSS baryon
acoustic peak, we obtain the best fit values of the EOS and two
major components' evolution. We find that the best fit EOS crossing
$-1$ in the near past $z\simeq 0.07$, the present best fit value of
$w_x(0)<-1$ and for this model the universe experiences the
acceleration at about $z\simeq 0.5$.
\end{abstract}


\section{Introduction}\label{I}
Observations of Cosmic Microwave Background (CMB)
anisotropies\cite{CMB}, high redshift type Ia supernovae\cite{Ia}
and the surveys of clusters of galaxies\cite{SDSS} indicate that an
exotic component with negative pressure dubbed dark energy dominates
the present universe. The most obvious candidate for this dark
energy is the cosmological constant $\Lambda$ with equation of state
($w_{\Lambda}=-1$), which is consistent with recent
observations\cite{CMB,WMAP06} in $2\sigma$ region. However, it
raises several theoretical difficulties\cite{Peebles03,Padmanabhan}.
This has lead to models for dark energy which evolves with time,
such as quintessence\cite{quintessence}, phantom\cite{phantom},
quintom\cite{quintom}, K-essence\cite{K-essence}, tachyonic
matter\cite{tachyon} and so on. For this kind of models, one can
design many kinds of potentials\cite{VS} and then study EOS for the
dark energy. Another way is to use a parameterization of the EOS to
fit the observational data, and then to reconstruct the potential
and the evolution of the universe\cite{reconstructure}. Various
parameterization of the EOS of dark energy have been presented and
investigated\cite{Corasaniti,Linder,Upadhye,Wang}.

If the universe has more than four dimensions, general relativity
should be extended from $4D$ to higher dimensions. One of such
extensions is the $5D$ Space-Time-Matter (STM)
theory\cite{Wesson,JMOverduin} in which our universe is a $4D$
hypersurface floating in a $5D$ Ricci-flat manifold. This theory is
supported by Campbell's theorem which states that any analytical
solution of the $ND$ Einstein equations can be embedded in a
$(N+1)D$ Ricci-flat manifold\cite{Campbell}. A class of cosmological
solutions of the STM theory is given by Liu and
Mashhoon\cite{LandMashhoon},the authors restudied the solutions and
pointed out that it can describe a bounce universe. It was shown
that dark energy models, similar as the 4D quintessence and phantom
ones, can also be constructed in this $5D$ cosmological solution in
which the scalar field is induced from the $5D$
vacuum\cite{ChangLiu,Liuetal}. The purpose of this paper is to use a
model-independent method to reconstructe a $5D$ cosmological model
and then  study the universe evolution and the EOS of the dark
energy which is constrained by recent observational data: the latest
observations of the $182$ Gold SNe Ia \cite{Riess06}, the 3-year
WMAP CMB shift parameter \cite{WMAP06,WangYun06} and the SDSS baryon
acoustic peak\cite{SDSS2}. The paper is organized as follows. In
Section 2, we briefly introduce the $5D$ Ricci-flat cosmological
solution and derive the densities for the two major components of
the universe. In Section 3, we will reconstruct the evolution of the
model from cosmological observations. Section 4 is a short
discussion.

\section{Dark energy in the $5D$ Model}\label{II}

The $5D$ cosmological model was described as
before\cite{LMW,LandMashhoon,Liu-b,zhang}. In this paper we consider
the case where the $4D$ induced matter $T^{\alpha \beta }$ is
composed of two components: dark matter $\rho _{m}$ and dark energy
$\rho _{x}$, which are assumed to be noninteracting. So we have
\begin{eqnarray}
\frac{3\left( \mu ^{2}+k\right) }{A^{2}} &=&\rho _{m}+\rho _{x},
\nonumber \\
\frac{2\mu \dot{\mu}}{A\dot{A}}+\frac{\mu ^{2}+k}{A^{2}}
&=&-p_{m}-p_{x},  \label{FRW-Eq}
\end{eqnarray}%
with
\begin{eqnarray}
\rho _{m} &=&\rho _{m0}A_{0}^{3}A^{-3},\quad p_{m}=0, \label{EOS-M} \\
p_{x} &=&w_{x}\rho _{x}.  \label{EOS-X}
\end{eqnarray}%
From Eqs. (\ref{FRW-Eq}) - (\ref{EOS-X}) and for $k=0$, we obtain
the EOS of the dark energy
\begin{equation}
w_{x}=\frac{p_{x}}{\rho _{x}}=-\frac{2\mu \dot{\mu}/\left(
A\dot{A}\right) +\mu ^{2}/A^{2}}{3\mu ^{2}A^{2}-\rho
_{m0}A_{0}^{3}A^{-3}},  \label{wx}
\end{equation}%
and the dimensionless density parameters
\begin{eqnarray}
\Omega _{m} &=&\frac{\rho _{m}}{\rho _{m}+\rho _{x}}=\frac{\rho
_{m0}A_{0}^{3}}{3\mu ^{2}A},  \label{omiga-M} \\
\Omega _{x} &=&1-\Omega _{m}.  \label{omiga-X}
\end{eqnarray}%
where $\rho _{m0}$ is the current values of dark matter density.

Consider Eq. (\ref{wx}) where $A$ is a function of $t$ and $y$.
However, on a given $y= constant$ hypersurface, $A$ becomes
$A=A(t)$, which means we consider a $4D$ supersurface embedded in a
flat $5D$ spacetime. As noticed before\cite{Xu-Recons,zhang}, the
term $\dot{\mu}/\dot{A}$ in (\ref{wx}) can now be rewritten as $d\mu
/dA$. Furthermore, we use the relation
\begin{equation}
A_{0}/A=1+z,  \label{factor}
\end{equation}%
as an ansatz\cite{Xu-Recons,zhang} and define $\mu _{0}^{2}/\mu
^{2}=f(z)$ (with $f(0)\equiv 1$), then we find that Eqs.
(\ref{wx})-(\ref{omiga-X})  can be expressed in term of the redshift
$z$ as
\begin{equation}
w_{x}=-\frac{1+(1+z)dlnf(z)/dz}{3(1-\Omega _{m})}, \label{wxz}
\end{equation}

\begin{equation}
\Omega_m=\Omega_{m_0}(1+z)f(z),  \label{Omegamz}
\end{equation}

\begin{equation}
\Omega_x=1-\Omega_m,  \label{Omegaxz}
\end{equation}

\begin{equation}
q=-\frac{1+z}{2}dlnf(z)/dz.  \label{qz}
\end{equation}%
where q is the deceleration parameter and $q<0$ meas our universe is
accelerating. Now we conclude that if the function $w_x$ is given,
the evolution of all the cosmic observable parameters in Eqs.
(\ref{wxz}) - (\ref{qz}) could be determined uniquely. Then we adopt
the parametrization of EOS as follows\cite{Linder,Eos-CP}
\begin{equation}
w_x(z)=w_0 + w_1 \frac{z}{1+z}\label{wz}
\end{equation}
From Eq. (\ref{wxz}) and Eq. (\ref{wz}), we can obtain the function
$f(z)$
\begin{equation}
f(z)=\frac{1}{(1+z)\left[\Omega_{m0}+(1-\Omega_{m0})(1+z)^{3w_0+3w_1}\exp(-\frac{3w_1
z}{1+z})\right]}.\label{fz3}
\end{equation}
In the next section, we will use the recent observational data to
find the best fit parameter ($w0,w1,\Omega_{m0}$).

\section{The best fit parameters from cosmological observations}\label{III}
In a flat universe with Eq. (\ref{wz}), the Friedmann equation can
be expressed as
\begin{equation}\label{hz}
H^2 (z)=H_0^2E(z)^2=H_0^2 [ \Omega_{0m} (1+z)^3
+(1-\Omega_{0m})(1+z)^{3(1+w_0+w_1)}e^{\frac{-3w_1 z}{(1+z)}}]
\end{equation}
Then the knowledge of $\Omega_{m0}$ and $H(z)$ is sufficient to
determine $w_x(z)$ with $H_0=72~{\rm \, km \cdot s}^{-1} \cdot {\rm
Mpc}^{-1} ~$\cite{HubbleConstant}. We use the maximum likelihood
method\cite{crossing} to constrain the parameters.

The Gold dataset compiled by Riess et. al is a set of supernova data
from various sources and contains 182 gold points by discarding all
SNe Ia with $z < 0.0233$ and all SNe Ia with quality='Silver' from
previously published data with 21 new points with $z> 1$ discovered
recently by the Hubble Space Telescope\cite{Riess06}. Theoretical
model parameters are determined by minimizing the quantity
\begin{equation}
\chi_{SNe}^2 (\Omega_{m0},w_0,w_{1})= \sum_{i=1}^N
\frac{(\mu_{obs}(z_i) - \mu_{th}(z_i))^2}{\sigma_{(obs; i)}^2}
\label{chi2}
\end{equation}where $N=182$ for Gold SNe Ia data,  $\sigma_{(obs; i)}^2$ are the
errors due to flux uncertainties, intrinsic dispersion of SNe Ia
absolute magnitude and peculiar velocity dispersion respectively.
These errors are assumed to be gaussian and uncorrelated. The
theoretical distance modulus is defined as
\begin{eqnarray}
\mu_{th}(z_i)&\equiv& m_{th}(z_i) - M \\\nonumber &=&5 \log_{10}
(D_L (z)) +5 \log_{10}(\frac{H_0^{-1}}{Mpc}) + 25 \label{mth}
\end{eqnarray}
where
\begin{equation}
D_L(z)=H_{0}d_L(z)=(1+z)\int_{0}^{z}\frac{H_{0}dz^{'}}{H(z^{'};\Omega_{m0},w_0,w_{1})}
\end{equation}
and $\mu_{obs}$ is given by supernovae dataset.

The shift parameter is defined as\cite{BondJR97}
\begin{eqnarray}
\bar{R}=\frac{l_1^{'TT}}{l_1^{TT}}= \frac{r_s}{r'_s}\frac{d'_A
(z'_{rec})}{d_A
(z_{rec})}=\frac{2}{\Omega_{m0}^{1/2}}\frac{q(\Omega_r',a_{rec})}{\int_0^{z}\frac{H_0
dz'}{H(z')}}
\end{eqnarray}
where $z_{rec}$ is the redshift of recombination, $r_s$ is the sound
horizon, $d_A(z_{rec})$ is the sound horizon angular diameter
distance, $q(\Omega_r',a_{rec})$ is the correction factor. For weak
dependence of $q(\Omega_r',a_{rec})$, the shift parameter is usually
expressed as
\begin{equation}
R=\Omega_{m0}^{1/2}{\int_0^{z}\frac{H_0
dz'}{H(z';\Omega_{m0},w_0,w_{1})}}
\end{equation}The R obtained from 3-year WMAP
data\cite{WMAP06,WangYun06} is
\begin{equation}
R=1.70\pm 0.03
\end{equation}
With the measurement of the R, we obtain the $\chi^2_{CMB}$
expressed as
\begin{equation}
\chi^2_{CMB}(\Omega_{m0},w_0,w_1)=\frac{(R(\Omega_{m0},w_0,w_1)-1.70)^2}{0.03^2}
\end{equation}

\begin{figure}[tpb]
\centerline{\psfig{file=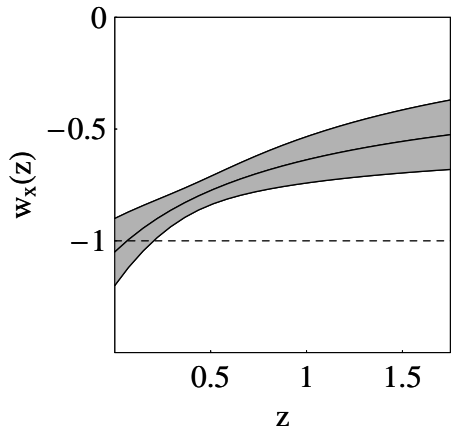,width=6cm}} \vspace*{8pt}
\caption{The best fits of $w_x(z)$ with $1\sigma$ errors (shaded
region).} \label{w-z}
\end{figure}

\begin{figure}[tpb]
\centerline{\psfig{file=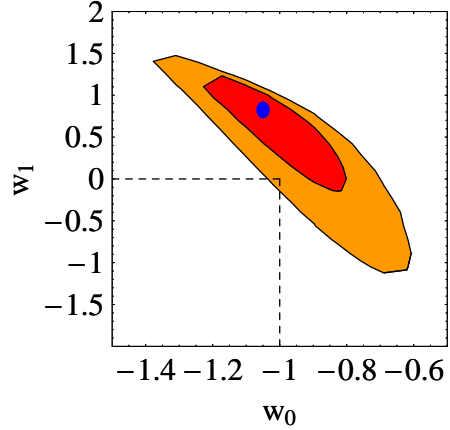,width=6cm}} \vspace*{8pt}
\caption{The contours show 2-D marginalized $1\sigma$ and $2\sigma$
confidence limits in the ($w_{0}$, $w_1$) plane} \label{w0-w1}
\end{figure}

The size of Baryon Acoustic Oscillation (BAO) is found by Eisenstein
et al\cite{SDSS2} by using a large spectroscopic sample of luminous
red galaxy from SDSS and obtained a parameter $A$ which does not
depend on dark energy directly models and can be expressed as
\begin{equation}
  A=\Omega_{m0}^{1/2}
E(z_{BAO})^{-1/3}[\frac{1}{z_{BAO}}\int_0^{z}\frac{dz'}{E(z';\Omega_{m0},w_0,w_{1})}]^{2/3}
\end{equation}
where $z_{BAO}=0.35$ and $A=0.469\pm0.017$. We can minimize the
$\chi^2_{BAO}$ defined as\cite{AlamU06}

\begin{equation}\label{BAO}
\chi^2_{BAO}(\Omega_{m0},w_0,w_1)=\frac{(A(\Omega_{m0},w_0,w_1)-0.469)^2}{0.017^2}
\end{equation}
To break the degeneracy of the observational data and find the best
fit parameters, we combine these datasets to minimize the total
lilkelihood $\chi^2_{total}$\cite{PDL}

\begin{eqnarray}
\chi^2_{total}=\chi^2_{SNe}+\chi^2_{CMB}+\chi^2_{BAO}
\end{eqnarray}
We obtain the best fit values $(\Omega_{m0},w_0,w_1)$ are (0.288,
-1.050, 0.824) and to identify the dependence of the best fit values
of the parameters, we set $\Omega_{m0}$ to be fixed when calculating
the confidence level of ($w0,w1$). The errors of the best fit
$w_x(z)$ are calculated using the covariance matrix\cite{AlamU04}
and shown in Fig.\ref{w-z}.  The corresponding $\chi^2$ contours in
parameters space (w0,w1) is shown in Fig.\ref{w0-w1}.

From Fig.\ref{w-z} we find that $w_x(z)$ is constrained in a narrow
space , the best fit $w_x(z)$ crosses $-1$ at about $z=0.07$ and at
present the best value of $w_x(0)<-1$, but in $1\sigma$ confidence
level we can't rule out the possibility $w_x(0)>-1$. Fig.\ref{w0-w1}
shows that a cosmological constant is ruled out in $1\sigma$
confidence level.

\begin{figure}[tpb]
\centerline{\psfig{file=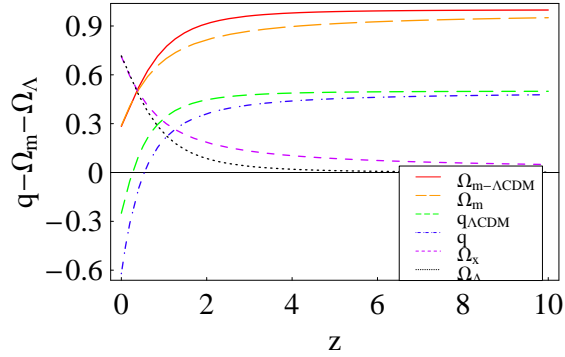,width=8cm}} \vspace*{8pt}
\caption{The evolution of q(z), $\Omega_m$, $\Omega_x$ and
q$_{\Lambda}$$_{CDM}$, $\Omega_{m-\Lambda CDM}$, $\Omega_{\Lambda}$
from 5D cosmological model and $\Lambda$CDM model respectly.}
\label{qmx}
\end{figure}

Using the function $f(z)$, the best fit values
$(\Omega_{m0},w_0,w_1)$, we obtain $\Omega_m$, $\Omega_x$,  the
deceleration parameter $q$ from Eq.(\ref{Omegamz})-(\ref{qz}) and
their evolution is plotted in Fig.\ref{qmx}. Fig.\ref{qmx} also
shows the evolution of q$_{\Lambda}$$_{CDM}$, $\Omega_{m-\Lambda
CDM}$, $\Omega_{\Lambda}$ in a $4D$ flat $\Lambda$CDM model with the
present $\Omega_{m0-\Lambda CDM}=0.283$ obtained from above
cosmological observations. We can see that the transition point from
decelerated expansion to accelerated expansion with $q=0$ is at
$z\simeq 0.5$ and it is earlier than the $\Lambda$CDM model. Our
universe experiences a expansion at present in a $4D$ supersurface
embedded in a $5D$ Ricci-flat spacetime or in $\Lambda$CDM model.

\section{Discussion}\label{IV}
Observations indicate that our universe now is dominated by two dark
components: dark energy and dark matter. The 5D cosmological
solution presented by Liu, Mashhoon and Wesson in\cite{LMW}
and\cite{LandMashhoon} contains two arbitrary functions $\mu (t)$
and $\nu (t)$, one of the two functions, $\mu(t)$, plays a similar
role as the potential $V(\phi)$ in the quintessence and phantom dark
energy models, which can easily change to another arbitrary function
$f(z)$. Thus, if the current values of the matter density parameter
$\Omega_{m0}$, $w0$ and $w1$ in the EOS are all known, this $f(z)$
could be determined uniquely. In this paper we mainly focus on the
constraints on this model from recent observational data: the $182$
Gold SNe Ia, the 3-year WMAP CMB shift parameter and the SDSS baryon
acoustic peak. Our results show that the recent observations allow
for a narrow variation of the dark energy EOS and the best fit
dynamical $w_x(z)$ crosses $-1$ in the recent past. Using the best
fit values $(\Omega_{m0},w0,w1)$, we have studied the evolution of
the dark matter density $\Omega_m$, the dark energy density
$\Omega_x$ and the deceleration parameter $q$ in a $4D$ supersurface
of $5D$ spacetime, which is similar to the $\Lambda$CDM model. In
the future, we hope that more and precision cosmological
observations could determine the key points of the evolution of our
universe, such as the transition point from deceleration to
acceleration, then distinguish the $5D$ cosmological model from
others.

\section{Acknowledgments}
This work was supported by NSF (10573003), NBRP (2003CB716300) of
P.R. China. The research of Lixin Xu was also supported in part by
DUT 893321 and NSF (10647110).

\end{document}